\documentstyle[pre, aps, twocolumn, psfig, epsf]{revtex}

\begin{document}
\twocolumn[\hsize\textwidth\columnwidth\hsize\csname @twocolumnfalse\endcsname

\title{Ground State Structure of Random Magnets}
\author{S. Bastea\cite{sb} and P. M. Duxbury\cite{pmd}}
\address{Department of Physics $\&$ Astronomy and
Center for Fundamental Materials Research,\\ Michigan State 
University, East Lansing, Michigan 48824-1116}
\maketitle
\begin{abstract}
Using exact optimization methods, we find all of the ground states 
of ($\pm h$) random-field Ising magnets (RFIM) and of dilute 
antiferromagnets in a field (DAFF). The degenerate ground states are 
usually composed of isolated clusters (two-level systems) embedded 
in a frozen background. We calculate the paramagnetic response 
(sublattice response) and the ground state entropy for the RFIM (DAFF) 
due to these clusters. In both two and three dimensions there is a broad 
regime in which these quantities are strictly positive, even at 
irrational values of $h/J$ ($J$ is the exchange constant).  
\end{abstract}

]

Disordered magnets \cite{young} provide a paradigm for
disordered systems in general, and they continue to be 
intensively
analyzed by a variety of methods.
However, due to metastability, 
conventional (e.g. Monte Carlo) analysis\cite{binder} of  
the equilibrium domain structure of random magnets is often 
unreliable,
especially at low temperatures. 
Since the ground state behavior is an important indicator of 
the low 
temperature behavior of most 
random magnets\cite{nattermann}, exact methods for ground state 
analysis are desirable.
 Fortunately, the {\it true} 
 ground states of random magnets can often be found  
using optimization methods. 

The relation between optimization and random magnets was
pointed out some time ago\cite{aura1,ogielski}.  However 
extensive use
of these methods is more recent, partially due to the
 availability of more efficient algorithms.  
An exact optimization procedure to find 
the  random field ground state was implemented by
Ogielski\cite{ogielski}.  More extensive analyses on larger 
system sizes have been
published recently\cite{aura2,usadel}.  These
  methods have also been extended to the analysis of the 
 ground state degeneracy of 
random magnets\cite{hartmann,hartmann1}.  
 Here we present, using a new algorithm, 
  a more precise analysis
of the ground state degeneracy and its consequences
 in  RFIM and DAFF in dimensions
d = 2,3 (square and cubic lattices).  We concentrate on 
the following three aspects of 
of these degenerate random magnets: 
(1) The degenerate domain
structure of RFIM  ground states (e.g. Fig. 1).
(2) The order parameter which couples to the
ground state degeneracy.
(3) The ground state entropy, and in particular the physical 
origin of its continuous and discontinuous parts.

We consider the random field Ising model (RFIM) with a 
binary random field,
\begin{equation}
H_{RFIM}=-J \sum_{(ij)} \sigma_i\sigma_j - \sum_i h_i\sigma_i
\end{equation}
where $h_i = \pm h$, with $h$ and $J$ positive and the
plus and minus random fields occur with equal probability.
We also analyze the dilute antiferromagnet in
a field (DAFF),
\begin{equation}
H_{DAFF}=  \sum_{(ij)} J x_{i} x_{j} \sigma_i\sigma_j - h 
\sum_i x_i\sigma_i
\end{equation}
where $p_i = p\delta(x_i-1) + (1-p) \delta(x_i)$ is the
probability that a site is present.  In both cases, we analyze
the ground state properties as a function of the ratio
$H = h/J$ on square and cubic lattices. 
 In the DAFF case there is the additional parameter
$p$ which we fix at $p=0.9$. 

The ground state critical behavior of random field magnets is 
 still not completely understood\cite{nattermann,aura2,banavar}.  
 At small $H$, large ferromagnetic domains
 are favored, while at large $H>H_*$, the spins freeze along
 the directions of the local field $h_i$.  In
 two dimensions there is no spontaneous magnetization for
 any finite $H$($H_c^{2d}=0$), though there is a rapidly 
growing
 ferromagnetic domain size  $l \sim exp (1/H^2) $ which can 
masquerade
 as a phase transition at $H\sim 1$.
 In three dimensions, there is a spontaneously magnetized state
 at small $H < H_c$.  Although the field theory analysis and
 early simulations suggested a continuous behavior
 in magnetization $m(H)$ as $H\rightarrow H_c^-$, 
 precise numerical work using exact optimization methods
 suggests a large jump in $m$ at $H_c$ (for the ($\pm h$) 
random field
 case  $H_c^{3d} = 2.21\pm 0.01$ and $\Delta m \sim 0.8$) 
\cite{aura2}.   
The DAFF was introduced as a possible experimental
realization of the RFIM\cite{fishman}, and an extensive 
literature
has developed from this observation\cite{belanger}. 
 The DAFF is
an antiferromagnet at small $H$, and at large $H>H_*$ 
all spins are polarized in the field direction.
 The DAFF {\it sublattice(staggered) magnetization}
is believed to be qualitatively similar to that
of the {\it magnetization} of the RFIM.
 Our calculations are for the ground
 state degeneracy in the non-trivial regime $0 < H < H_*$, 
where $H_* = 2d$
 ($d$ is dimension) is
 the field amplitude beyond which {\it all} spins follow the 
local field    
 direction and the ground state is non-degenerate 
 (for both RFIM and DAFF). 
 
 The ground state degeneracy of the RFIM has been intensively
 studied in one dimension\cite{derrida}.  There has also been 
an analysis
 on Cayley trees\cite{bruinsma}, and an interesting analysis 
 on the square lattice\cite{morgenstern}.
 The latter paper did not have the advantage of exact 
 optimization methods and missed some of the key features of
 the ground state degeneracy. Although the infinite range model
 misses entirely the degeneracy we find here\cite{bruinsma}, 
 the one dimensional
 and Cayley tree models have several qualitative similarities
 with our results.  More recently Hartmann\cite{hartmann1} has presented
 a low precision calculation of the ground state degeneracy of
 random magnets, though the physics we elucidate here was
 not discussed by him.
 
\begin{figure}
\centerline{\epsfxsize=3.2in\epsfbox{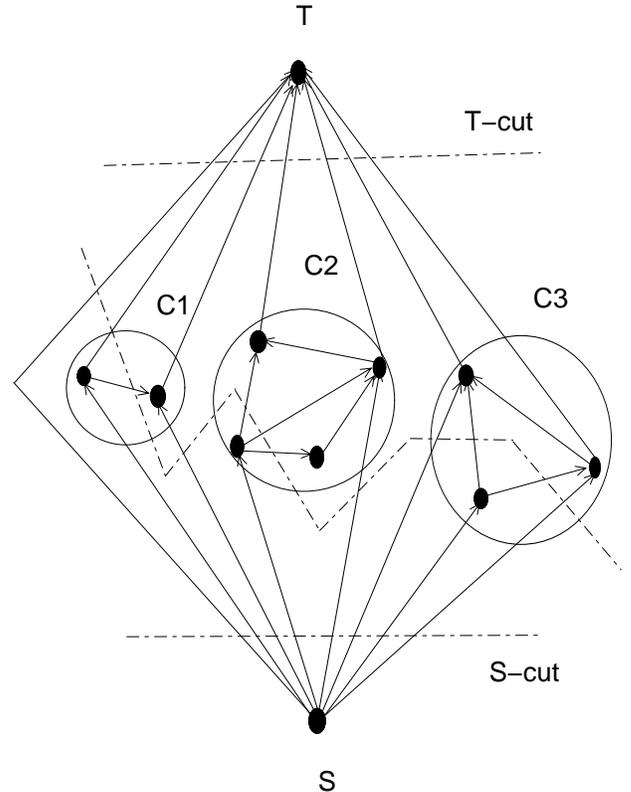}}
\caption{Typical supergraph as obtained using the algorithm. 
$S$ is the set of spins frozen up and $T$ the set of spins frozen down. 
$C_1$, $C_2$ and $C_3$ are independent clusters; they are made of 
 subclusters that are not independent of each other. The 
$S-cut$ is the ground state with all the independent clusters down, and 
the $T-cut$ is the one with all of them up. We also show a directed cut 
(ground state) in which part of each independent cluster is up/down.  }
\label{superg}
\end{figure}

In order to find the ground state of 
RFIM and DAFF, we use the mapping of these
problems to a flow problem in combinatorial optimization
(so called min-cut/max flow)\cite{comb}. 
 This algorithm also
gives the exact minimal energy surface in
random networks\cite{middleton,rieger}.
This has been known for some time\cite{ogielski}, however
improved algorithms  (push-relabel with global 
updates\cite{goldberg}) 
 now allow optimization of $100^3$ lattices in
a few minutes on a high end workstation. 
Our method relies on the concept of {\it residual graph} 
introduced
by the network flow algorithms \cite{goldberg}. The full 
residual graph 
of the equivalent
network flow problem holds the whole information about the 
ground state
structure. A naive search of the ground states - which is 
equivalent to 
finding the domains that can be flipped
without altering the energy (or the max-flow in the network 
flow terminology) 
- is exponential. Instead we generate a supergraph which
shows how the domains are related to each other, i.e. which
domains can be flipped independently. It turns out that
many of the domains are independent and the exponential search
is reduced to the few remaining dependent domains, and 
we search over these remaining domains.  We show that the 
problem is equivalent to finding all the directed cuts in a 
directed graph with single arcs and no cycles. A typical 
supergraph is shown in Fig. \ref{superg}.
It is easy to see how the structure of the supergraph makes 
such a search effective, as it reduces it to searches over much 
smaller independent graphs. Details of the method will be presented 
elsewhere \cite{bastea1}.

\begin{figure}
\centerline{\psfig{file=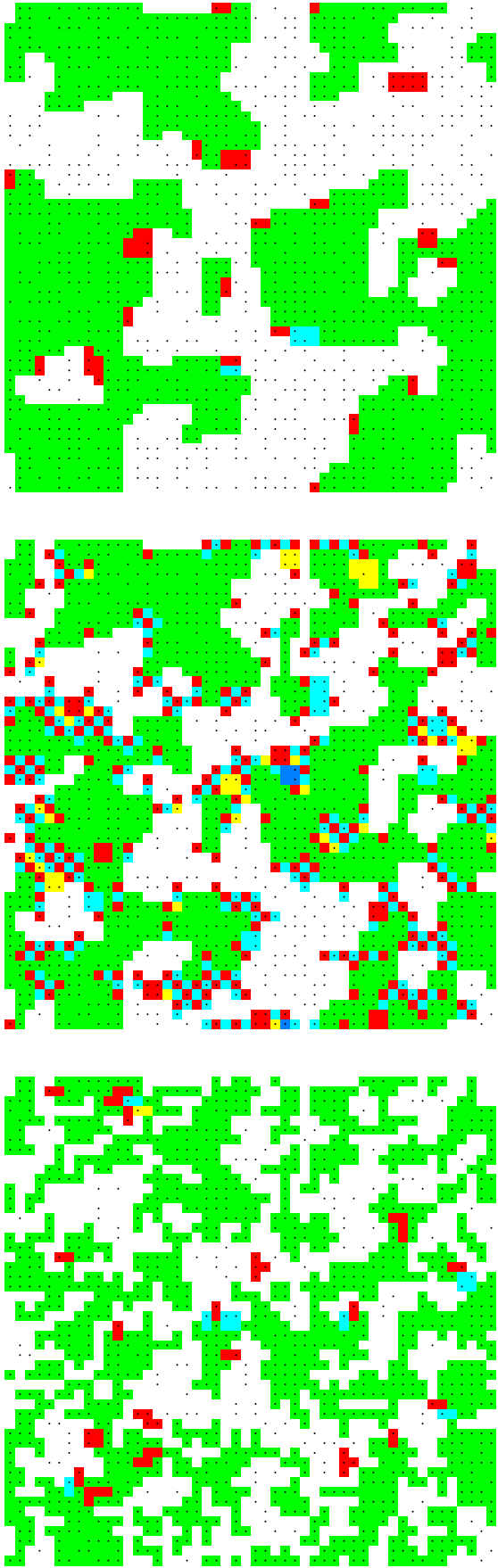,width=2.9truein,angle=180}}
\caption{Typical ground states of the $d=2$ random field
ising model (Eq. 1) with $H= h/J = 3/2$ (top), $H=2$ (middle)
$H=5/2$ (bottom) (system size $50\times 50$).  Green indicates 
an up spin, 
white a down spin, and
the other colors indicate spin clusters that can be flipped up 
or
down (but not all independently of each other)
 without changing the ground state energy. Black dots indicate 
the
sites where the random field favors the up spin orientation.}
\label{gs}
\end{figure}

\begin{figure}
\centerline{\psfig{file=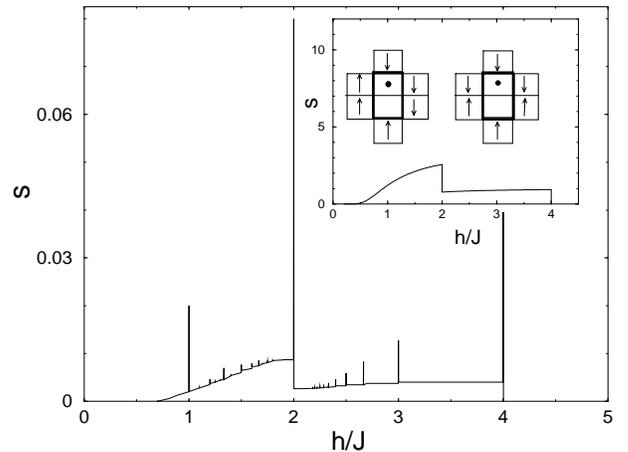,width=3.2truein,angle=-90}}
\caption{ The ground state entropy of the 2-d random field 
Ising
model (RFIM) as a function of $h/J = H$.  The inset shows the
smallest two-level-systems (TLS's) at irrational $H$
and an estimate of the entropy (from Eq. (3)) produced by them.
 ``+'' indicates an up spin
``-'' indicates a down spin and a dot indicates where the
local random field favors the ``+'' spin direction. The system 
sizes used 
were from $10\times 10$ to $130\times 130$ and the entropy was 
found as 
the slope of the line $\langle\ln D\rangle.vs.N$, where D is 
the degeneracy, 
N is the system size (total number of spins) and the average is 
over the 
disorder (1000 samples were used).}
\label{tls}
\end{figure}

A typical ground-state domain structure of the  
two-dimensional $\pm h$ RFIM
is presented in Figs. \ref{gs}. Here green domains are
fixed in the direction of the positive fields (dots), while
white domains are fixed in the opposite direction. Domains 
of any other color can be flipped without changing
the ground state energy (note that not all can be flipped
independently, but the dependent domains are organized in 
clusters 
that can also be flipped as a whole). These domains produce a 
finite
ground state entropy. Surprisingly, domains that can be flipped 
 in the ground state exist {\bf even for irrational $H$ = 
$h/J$}.
Note that these domains occur at the {\it interfaces}
between the up-spin and down-spin domains of the RFIM
ground state. The degenerate clusters at irrational $H$
 have {\it zero field energy}
{\bf and} {\it the same exchange energy} in both the up and
down states of the cluster.
 For the RFIM on the square lattice, the lowest order
 degenerate clusters of this sort are indicated in the inset to 
Fig. \ref{tls}.
 The number of these clusters (or two-level-systems (TLS)),
  $n_{TLS}$, can be estimated using simple 
arguments:
\begin{equation}
n_{TLS}\propto p_{TLS}\frac{L^2}{l}
\end{equation}
where $L$ is the linear size of the system, $l\propto 
exp[(1/H)^2]$ is the 
typical size of the ordered domains \cite{nattermann} and 
$L^2/l$ is the total length of interface between up and down 
spin domains
in the system. $p_{TLS}$ is
the probability of occurrence of a TLS at a given interface 
site.
 $p_{TLS}=p_n/4$, 
where $1/4$ is the 
probability of occurrence of the up-down pair of fields and 
$p_n$ is the 
probability to have this pair surrounded in the ground state 
by frozen spins with the appropriate 
configurations.  The entropy density is then
$s\propto p_n exp[-(1/H^2)]$ for $H< 4$ and $0$ for $H> 4$. 
 $p_n$ is discontinuous at $H=2$, because the
 dominant TLS's for $H<2$ are different than those for $H>2$
(for example there are twice as many spin configurations that 
lead
to a TLS below than above $H=2$).
 If we take the observed jump $\sim 3.3$, then the above
argument leads to the curve given in the inset to Fig. 
\ref{tls},
which is very close in form to the continuous part of the
entropy presented in Fig. \ref{tls}.
 
 The series of sharp peaks
occurring at rational values of $H$ are due
to additional degeneracy occurring when 
clusters have same  value for 
{\it the field energy} {\bf plus} {\it the exchange energy}
in either the up or down states. These peaks can only occur
at {\it rational} values of $H$ and the cluster geometries
which contribute at each rational are different.  Naturally
high order rationals correspond to complex clusters and
 have greatly reduced degeneracy.  
It has been  suggested before \cite{morgenstern} that
 for the 2d RFIM, the highest degeneracies 
  occur at rationals $H_n=2+2/n$ for $2<H<4$, with $n=1,2...$.  
Using 
 our algorithm we checked this idea by considering $n=1,..11$
 and all rationals with denominators $2,3,4,5,6$.  We find that
 those with $H_n=2+2/n$ are indeed dominant \cite{only}, see 
Fig. \ref{tls}, 
and there is a similar sequence at fields $4+2/n$ in 3d (the
 zoology of the clusters leading to the dominant peaks is
 straightforward though tedious to enumerate). In the
 regime $0<H<2$ the 2-d RFIM ground-state entropy has spikes
 at a large number of rationals (see Fig. \ref{gs}).  These
 features are quite similar to those found in one dimension
 (see Fig. 4 of \cite{derrida}) and on  Cayley trees 
 (see Fig. 4 of \cite{bruinsma} ).
 We have also done a preliminary analysis of 
 the 3-d RFIM and the 2-d and 3-d DAFF ground states.
 In general we find that the RFIM and DAFF magnets in
 2-d and 3-d are massively degenerate in the regime $H_c<H<H_*$
 and that their ground state entropy is finite even at 
 irrational $H$.  

\begin{figure}
\centerline{\psfig{file=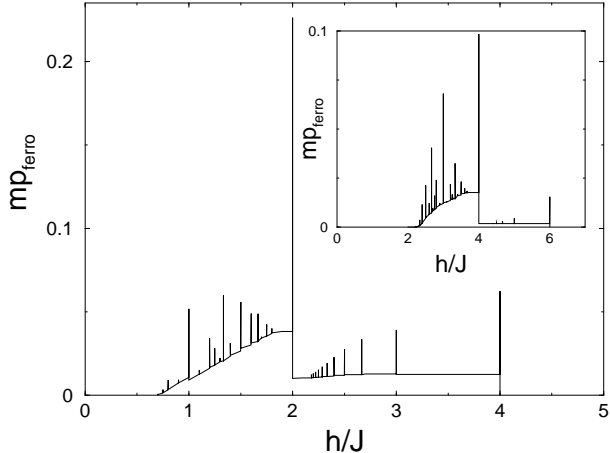,width=3.2truein,angle=-90}}
\caption{The order parameter for the ground state
paramagnetism ($mp_{ferro}$) of the RFIM on square 
and cubic(inset) lattices. System size was $200\times 200$ for 
the square
and $40\times 40\times 40$ to $60\times 60\times 60$ for the 
cubic lattice 
(1000 samples were used).}
\label{rfim}
\end{figure}

In the regime $H_c<H\le H_*$, we can consider the ground
state to be composed of a frozen background in which is 
embedded a
set of largely non-interacting free {\it superspins}
(corresponding to each independent cluster).  The ground
state of these magnets thus can be considered to contain a 
large number of
magnetic two-level-systems\cite{copper}.  There is a 
{\it paramagnetic response} at low temperatures for both
the RFIM and DAFF in this regime. 
The natural ground state order parameter for the
paramagnetic response in the regime $H_c<H<H_*$
is the magnetization $mp_{ferro}$ for
the RFIM and the staggered magnetization $mp_{staggered}$
for the DAFF.  It is straightforward to calculate these
order parameters using the exact optimization algorithm, either
by applying an appropriate infinitesimal field or by polarizing
all of the degenerate domains in a given orientation (we do the 
latter). 
 The results
for the RFIM are presented in Fig. \ref{rfim} for both square 
and cubic (inset)
lattices.

It is seen that the basic
features of the ground state degeneracy (Fig. \ref{tls}) are 
reflected in the
ground state paramagnetic magnetization.    
 In 3-d, the entropy remains
 zero at low $H$ (at least for H irrational), reflecting
the ferromagnetic state for $H<H_c \sim 2.21$\cite{aura2}.   
For
$H>H_* = 6$ (in 3-d), the spins are aligned with the random 
field, and the
ground state is non-degenerate.   Note that in addition to the
paramagnetic magnetization, there is a spontaneous 
magnetization
$ms_{ferro}$ for $H<H_c$.  In experiments in which the 
ferromagnetic field
is swept to produce a magnetization loop, the measured
zero field magnetization is the sum i.e. $m0_{ferro} = 
ms_{ferro}
+ mp_{ferro}$.  Thus there is a finite equilibrium 
magnetization jump  
at zero temperature even for $H>H_c$.  Of course
$mp_{ferro} = 0$ for $T>0$, but the effects of the ground state
degeneracy should be reflected in magnetization anomalies and 
a Curie Law in the susceptibility at low temperatures.   

\begin{figure}
\centerline{\psfig{file=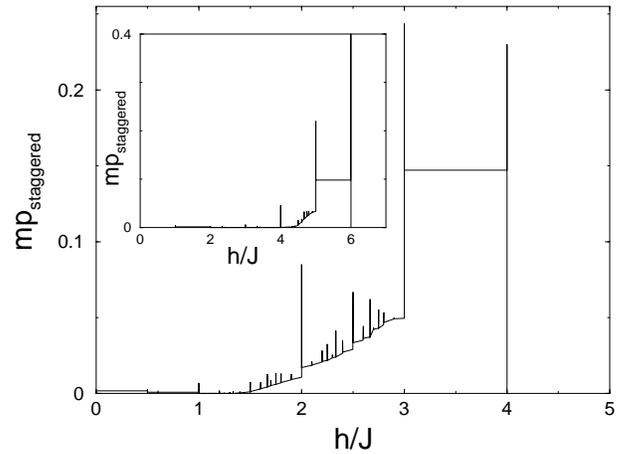,width=3.2truein,angle=-90}}
\caption{The order parameter for the ground state
paramagnetism ($mp_{staggered}$) of the DAFF  
on square and cubic (inset) lattices. The same system sizes and 
number 
of samples as for the RFIM were used.}
\label{daff}
\end{figure}

Calculations of the paramagnetic order parameter for the 
DAFF, $mp_{staggered}$,
for square and cubic lattices is presented in
Fig. \ref{daff}.
Qualitatively the situation is similar to 
that in
the RFIM.  There is a strong sublattice paramagnetic
response for all
$H_c< H < H_*$, with spikes at certain rational values.  These
figures are for a DAFF with dilution $p=0.9$, but only
the details change as $p$ is varied.  In the regime
$H<H_c$ there is a spontaneous staggered magnetization, and
low temperature measurements (such as neutron scattering
and NMR) should be influenced by both
the ``staggered paramagnetic'' response and the 
spontaneous staggered magnetization. We also note that the existence 
of the additional order parameter $mp_{ferro}$ in the case of the $\pm h$ RFIM 
suggests that the $\pm h$ RFIM and the Gaussian RFIM may not be in the 
same universality class.
  
We have described two developments
in the analysis of random magnets.\\
(i) Using optimization methods it is possible to
efficiently calculate the ground state structure of RFIM 
and DAFF magnets (see Fig. 2).  
(ii) The ($\pm h$) RFIM and the DAFF magnets have a 
spontaneously ordered state for $H<H_c$, a massively degenerate 
ground state in the regime $H_c\le H \le H_*$ and
a non-degenerate ground state for $H>H_*$.
In the degenerate regime 
there is a strictly positive paramagnetic response and ground state 
entropy even at irrational $H$, with additional degeneracies at 
rational $H$ (see Figs. 2-4).\\

We thank Michael F. Thorpe and David P. Belanger for discussions.
This work was supported by the DOE under contract
DE-FG02-90ER45418.


\begin{references}

\bibitem[\dag]{sb}Electronic address: bastea@pa.msu.edu
\bibitem[\star]{pmd}Electronic address: duxbury@pa.msu.edu
\bibitem{young}For a review see for example, {\it Spin Glasses 
and Random Fields}, edited by A. P. Young (World Scientific, 
Singapore, 
1997).
\bibitem{binder}K. Binder and D. W. Heermann, {\it Monte Carlo 
Simulation in  Statistical Physics} (Springer-Verlag, 1992).
\bibitem{nattermann} T. Nattermann and J. Villain,
Phase Transitions {\bf 11}, 5 (1988);
T. Nattermann, {\it Theory of the Random 
Field Ising Model}, in [1].
\bibitem{aura1} J.C. Angl\`es d'Auriac, M. Preissmann and R. 
Ramal,
J. de Physique Lett. {\bf 46}, L173 (1985).
\bibitem{ogielski}A. T. Ogielski, Phys. Rev. Lett. {\bf 57}, 
1251, (1986).
\bibitem{aura2} J.C. Angl\`es d'Auriac and N. Sourlas, 
Europhys. Lett.
{\bf 39}, 473 (1997).
\bibitem{usadel}J. Esser, U. Nowak and K. D. Usadel, Phys. Rev. 
B {\bf 55}, 5866, (1997).
\bibitem{hartmann}A. Hartmann and K. D. Usadel, Physica A {\bf 
214}, 141 (1995).
\bibitem{hartmann1}A. Hartmann, Physica A {\bf 248}, 1 (1998).
\bibitem{banavar} M.R. Swift, A. Maritan, M. Cieplak and J. 
Banavar,
J. Phys. {\bf A27}, 1525 (1994).
\bibitem{fishman} S. Fishman and A. Aharony, J. Phys. {C12}, 
L729 (1979);
J.L. Cardy, Phys. Rev. {\bf B29}, 505 (1984).
\bibitem{belanger} D.P. Belanger, {\it Experiments on the 
Random Field Ising Model} in [1].
\bibitem{derrida} B. Derrida, J. Vannimenus and Y. Pomeau,
J. Phys. {\bf C11}, 4749 (1978).
\bibitem{bruinsma} R. Bruinsma, Phys. Rev. {\bf B30} (1984).
\bibitem{morgenstern} I. Morgenstern, K. Binder and R.M. 
Hornreich,
Phys. Rev. {\bf B23}, 287 (1981).
\bibitem{comb}J. H. van Lint and R. M. Wilson, {\it A Course in 
Combinatorics}, Chapter 7 (Cambridge University Press, 1996); 
A. V. Goldberg, E. Tardos and R. E. Tarjan, 
Network Flow  Algorithms, in {\it Paths, Flows, and 
VLSI-Layout} 
(Springer-Verlag, 1990).
\bibitem{middleton}A. A. Middleton, Phys. Rev. E {\bf 52}, 
R3337, (1995); M. J. Alava and P. M. Duxbury, Phys. Rev. B 
{\bf 54}, 14990, (1996).
\bibitem{rieger}H. Rieger, Frustrated Systems: Ground State Properties via
Combinatorial Optimization, in {\it Lecture Notes in Physics: Advances in 
Computer Simulation}, edited by J. Kertesz and I. Kondor, (Springer-Verlag, 
1998).
\bibitem{goldberg}A. V. Goldberg and R. E. Tarjan, J. Assoc. 
Comput. Mach.  {\bf 35}, 920, (1988).
\bibitem{bastea1}S. Bastea, {\it A Degeneracy Algorithm for Random Magnets 
and Other Systems}, cond-mat/9807260, submitted to Phys. Rev. E.
\bibitem{only} However they are not the only ones, but all the 
others are very strongly supressed.
\bibitem{copper} S. N. Coppersmith, Phys. Rev. Lett. {\bf 67}, 
2315, (1991).

\end{references}
\end{document}